\documentclass[aip,reprint]{revtex4-1}

\usepackage[utf8]{inputenc}
\usepackage[english]{babel}
\usepackage[T1]{fontenc}
\usepackage{csquotes}

\usepackage{amssymb,amsmath,amsthm}
\usepackage{mathrsfs,mathtools}

\usepackage{bm,color}

\newcommand{\de}{\partial}
\newcommand{\vc}[1]{\boldsymbol{#1}}  
\newcommand{\vt}[1]{\mathsf{#1}}  
\newcommand{\tsp}{\mathsf{T}}  

\newcommand{\tr}{\operatorname{\mathrm{tr}}}
\newcommand{\dvg}{\operatorname{\mathrm{div}}}

\newcommand{\DD}{\vt{D}}

\newcommand{\F}{\vt{F}}
\newcommand{\Fr}{\vt{F}_\mathrm{R}}
\newcommand{\G}{{\vt{G}}}

\newcommand{\I}{{\vt{I}}}

\newcommand{\Adv}[1]{\mathcal{D}_{#1}}
\newcommand{\PL}{{\boldsymbol{\varphi}}}

\newcommand{\revision}[1]{\textcolor{black}{#1}}

\newcommand{\taur}{\tau_\mathrm{r}}
\newcommand{\T}{{\vt{T}}}
\newcommand{\Tel}{\T_\mathrm{el}}

\newcommand{\gammar}{\gamma_\mathrm{R}}
\newcommand{\gd}{\dot{\gamma}}

\newcommand{\gammaj}{\gamma_\mathrm{J}}

\newcommand{\etaeff}{\eta_\mathrm{eff}}

\newcommand{\psieff}{\Psi_\mathrm{eff}}

\newcommand{\Tvi}{\T_\mathrm{vi}}

\newcommand{\Bel}{\vt{B}_\mathrm{el}}
\newcommand{\Cel}{\vt{C}_\mathrm{el}}
\newcommand{\Fel}{\vt{F}_\mathrm{el}}
\newcommand{\cauchy}{\boldsymbol{\sigma}}

\newcommand{\De}{\mathit{De}}
\newcommand{\Wi}{\mathit{Wi}}

\theoremstyle{definition}

\makeatletter
\def\@email#1#2{%
 \endgroup
 \patchcmd{\titleblock@produce}
  {\frontmatter@RRAPformat}
  {\frontmatter@RRAPformat{\produce@RRAP{*#1\href{mailto:#2}{#2}}}\frontmatter@RRAPformat}
  {}{}
}%
\makeatother

\begin{document}

\title[Evolution of local relaxed states and the modelling of viscoelastic fluids]{Evolution of local relaxed states and the modelling of\\ viscoelastic fluids}

\author{Muhanna Ali H \surname{Alrashdi}}
\altaffiliation[Present affiliation: ]{College of Science, Department of Mathematics, University of Hail}

\author{Giulio Giuseppe \surname{Giusteri}}
\email{giulio.giusteri@unipd.it}

\affiliation{Dipartimento di Matematica ``Tullio Levi-Civita'', Universit\`a degli Studi di Padova\\ via Trieste 63, 35131, Padova, Italy}

\date{\today}

\begin{abstract}%
We introduce a class of continuum mechanical models aimed at describing the behaviour of viscoelastic fluids by incorporating concepts originated in the theory of solid plasticity.
Within this class, even a simple model with constant material parameters is able to qualitatively reproduce a number of experimental observations in both simple shear and extensional flows, including linear viscoelastic properties, the rate dependence of steady-state material functions, the stress overshoot in incipient shear flows, and the difference in shear and extensional rheological curves.
Furthermore, by allowing the relaxation time of the model to depend on the total strain, we can reproduce some experimental observations of the non-attainability of steady flows in uniaxial extension, and link this to a concept of polymeric jamming or effective solidification.
Remarkably, this modelling framework helps in understanding the interplay between different mechanisms that may compete in determining the rheology of non-Newtonian materials.
\end{abstract}

\maketitle

\section{Introduction}

Rheology deals with the way materials deform when forces are applied to them. The subject of its studies comprises both fluid-like and solid-like behaviours, that can appear in combination when dealing with complex fluids. At opposite ends of the viscoelastic spectrum we find Newtonian viscous fluids and elastic solids. The former resist shear by dissipating energy and can strain indefinitely under an applied stress, the latter can store energy and balance the applied stress to reach a static configuration, always remembering their original shape.
To be able to describe intermediate material responses, we need to address the evolution of the elastically relaxed shape, as well as that of the actual deformation.

Viscoelastic models in continuum mechanics go all the way back to Maxwell and reached a fundamental turning point in their systematic treatment with the work of Oldroyd~\cite{oldroyd_1950,oldroyd_1958}. In the approach of differential models, the flow equations are typically coupled to an evolution equation for either the elastic stress itself or a conformation tensor. The time-dependence of the elastically relaxed shape remains implicit in these relations, whereas a more direct description of it could help linking the mathematical model to its mechanical origin.
Among the several important contributions that are by now textbook material \cite{Bird_1987,Phan-Thien_2013}, we would like to highlight the models by Giesekus~\cite{giesekus_1966,giesekus_1982} and by Phan-Thien \& Tanner~\cite{thien_1977} for their tensorial structure and usefulness in fitting experimental data. 
Very informative reviews of Oldroyd's influence on complex fluids modelling and open issues can be found in recent articles by Beris~\cite{Beris_2021} and Renardy \& Thomases \cite{Renardy_2021}.

The large quantity of complex fluids and mixtures that are relevant for engineering applications requires the constant development of new mathematical models.
We believe that focusing on the evolution of the elastically relaxed microsctructure as a primary variable can help in the construction of models that, on one hand, can be more easily related with measurable effects and, on the other hand, have a tensorial structure suitable for large-scale simulations of generic flows. Hence, the starting point of our approach is to obtain the evolution of the elastic stress as a result of the evolution of the current deformation gradient and of a tensorial description of the local relaxed state. 
The evolution of the former quantity is fully determined by the continuum kinematics, while the structure of the equation for the relaxed state is similar to the flow rules that are postulated in plasticity theory~\cite{Gurtin_2010}.
This perspective was introduced in a seminal work by Rajagopal \& Srinivasa~\cite{Rajagopal_2000} and, thanks to its generality, has been applied in several contexts, such as the mechanics of asphalt~\cite{Krishnan_2005}, the description of thixotropic materials~\cite{deSouzaMendes_2012}, blood rheology~\cite{Apostolidis_2015,Giannokostas_2020}, the theory of shape-memory and crystallizing polymers~\cite{Yarali_2020,Sreejith_2023}, and food rheology~\cite{vanderSman_2024}.

Within this line of research, our work has some distinctive features. First of all, we do not make use of the concept of natural configuration as a virtual stress-free deformation. Indeed, it is by now well understood that it is not necessary for the tensor field that locally describes the elastically relaxed material state to be the gradient of any mapping. Secondly, our constitutive choices produce a thermodynamically consistent model, but they are based on considerations of a purely mechanical nature on the evolution of the relaxed state and on the stress response rather than on variational and thermodynamical arguments.
Within our framework, viscoelastic fluids emerge as an interpolation between purely viscous fluids and solids, controlled precisely by the characteristic time of the plastic evolution.
This is a feature not always present in viscoelastic models, as pointed out by Snoeijer \emph{et al.}\ \cite{Snoeijer_2020} in a recent review.
To achieve our goals, we use well-established concepts from solid plasticity to address the modelling of viscoelastic fluids and show that one can create a fruitful synergy between the fluid and solid perspectives when modelling complex materials. We offer an alternative to Oldroyd's approach by taking more explicitly into account plastic effects. 
We present a tensorial model that features the same number of parameters as the upper-convected Maxwell model, but displays improved linear viscoelastic properties akin to those of the Giesekus model and, most importantly, is able to capture rate-dependent effects in transient shear flows as well as flow-type dependence of the rheological curves. We stress that these desirable features emerge as a consequence of the mechanical framework we employ and are not added to our model by design. This marks a definite conceptual advantage in comparison to more complex multi-parameter models.

A second important aspect is the use of logarithmic relations between the elastic stress and the strain measure, akin to the Hencky strain approach. 
Importance and advantages of logarithmic strains have been discussed in the literature by several authors \cite{Xiao_2000,Xiao_2004,Neff_2016,Prusa_2020} and in connection with viscoelastic fluid models by Pr{\r{u}}{\v{s}}a \& Rajagopal~\cite{Prusa_2021}, albeit in the context of implicit constitutive relations which is quite different from our approach.
Our motivation for this choice comes from a very specific and fundamental aspect. Stresses represent forces that drive the material deformation and, as such, they naturally belong to a linear space in which it is meaningful to take the sum of different contributions to form the total stress. On the other hand, deformation gradients belong to the multiplicative group of linear transformations and the basic measures of strain---the Cauchy--Green tensors---are Riemannian metric tensors. For all these objects, the algebraic operation of sum does not represent any physically relevant operation and we wish to avoid it in developing constitutive relations. The matrix logarithm provides a mapping from the multiplicative group of deformation gradients to its Lie algebra, which is indeed a linear space suitable to host the description of stresses.
This argument can be summarized by saying that the use of mathematical representations of physical quantities must respect their mechanical meaning. I our view, the fundamental role of logarithmic strains is to guarantee such consistency.

The use of scalar logarithmic strains in continuum mechanics has a long history~\cite{Neff_2016} and the tensorial version proposed by Hencky~\cite{Hencky_1933} has become a fundamental object in textbooks on nonlinear elasticity~\cite{Gurtin_2010}. Hencky's studies were also relevant in foundational developments in theoretical rheology~\cite{Tanner_2003}. One of the original motivations for the use of logarithmic stress-strain relations by Hencky was that such models prevent the total compression of the material, which should be avoided due to the unphysical interpenetration of matter that would occur. A second mechanical issue in which logarithmic relations are helpful is a proper separation of the stress response to spherical volumetric changes and shear deformation~\cite{Xiao_2004}, which often display very different stiffness and somewhat independent origins. Furthermore, the Hencky strain has been used with the goal of identifying the direction (in a tensorial space) along which the elastic energy can be minimized in an optimal way, thereby selecting a more physical stress tensor~\cite{Neff_2016}. Our mathematical arguments given above attempt at a rigorous translation into the constitutive model of all these desirable features. The use of logarithmic strains has recently found application in describing the elastic contribution to jamming of granular matter~\cite{Luding_2021} and dense suspensions~\cite{Giusteri_2021}.

Within the general setting that we propose, we first specialize our treatment to the basic case in which material parameters are assumed constant. This can of course be relaxed to better represent real fluids. Nevertheless, we can already capture nontrivial qualitative features of viscoelastic flows. 
We are able to obtain a dependence of rheological curves on the shear rate and even on the flow type without any such dependence in the material parameters.
Indeed, those effects turn out to be a result of how the different flows affect the elastic part of the stress, with particular reference to the dynamics of the principal stress directions.

The paper is organized as follows. In Section \ref{sec:constitutive}, we introduce the basic evolution equations and a new class of viscoelastic constitutive models. 
Section~\ref{sec:energy} is devoted to finding the evolution equation for the elastic strain (akin to a conformation tensor) implied by our model and to the analysis of the energy balance, that highlights the presence of two distinct dissipative effects.
Focusing on the special case of a model with constant coefficients, we present in Section~\ref{sec:small_amplitude} analytical results about small-amplitude oscillatory flows,
while stress growth and stress relaxation predictions are reported in Section~\ref{sec:stress_growth} and \ref{sec:stress_relax}, respectively. To give a first example of the flexibility of our approach, we introduce, in Section~\ref{sec:jamming}, a model that captures a strain-induced fluid-solid transition sometimes observed in transient uniaxial extension measurements.
Moreover, in Section~\ref{sec:developing}, we exploit further the possibility of using non-constant relaxation times to reproduce the rheological data of wormlike micellar solutions and comment on the limitations of our model with respect to the description of materials with multiple relaxation times.
Dimensionless numbers that can be associated with the proposed model are discussed in Section~\ref{sec:dimensionless}.
In the final Section~\ref{sec:conclusions}, we conclude by summarizing our main results and addressing further research directions.

We aimed at giving a concise presentation of a general modelling strategy for viscoelastic materials and of a few paradigmatic examples of its effectiveness in reproducing the qualitative features of experimental measurements. 
We kept technicalities to a minimum while presenting rigorous computations to help the reader grasp the key features of our approach.
Several extensions of this work can be foreseen, especially in the direction of including multiple relaxation times and finite extensibility effects.
What we wish to stress once more is the importance of taking into account an evolution equation for tensorial descriptors of the local elastically-relaxed state as a central part of constitutive models.

\section{Flow equations, kinematics, and constitutive relations}\label{sec:constitutive}

In this section, we introduce the basic evolution equations for a general continuum in the Eulerian setting. We denote by $\rho$ the mass density, that we assume constant and uniform, and by $\vc u(\vc x,t)$ the velocity field at point $\vc x$ and time $t$.
To be able to follow the material deformation, we introduce the placement $\PL(\vc X,t)$ that gives the position at time $t$ of a material point labelled with $\vc X$ and solves the nonlinear equation $\de_t \PL(\vc X,t)=\vc u(\PL(\vc X,t),t)$.
We define the Eulerian deformation gradient $\F(\vc x,t):=\nabla_{\vc X}\PL\vert_{\vc X=\hat{\PL}(\vc x,t)}$ with $\hat{\PL}$ the spatial inverse of $\PL$, such that $\PL(\hat{\PL}(\vc x,t),t)=\vc x$.
For the sake of clarity, we use the following notation for the material time derivative:
\[
\Adv{\vc u}:=\frac{\de}{\de t}+\vc u\cdot\nabla.
\]

The balance of linear momentum gives rise to the evolution equation~\cite{Gurtin_2010} 
\begin{equation}\label{eq:motion}
\rho\Adv{\vc u}\vc u=\dvg\cauchy
\end{equation}
for the velocity field, driven by the Cauchy stress tensor $\cauchy$. We neglect external forces for simplicity.
Incompressiblity is expressed by the constraint $\dvg\vc u=0$.

\begin{figure}[t]
\centering
\includegraphics[width=0.9\columnwidth]{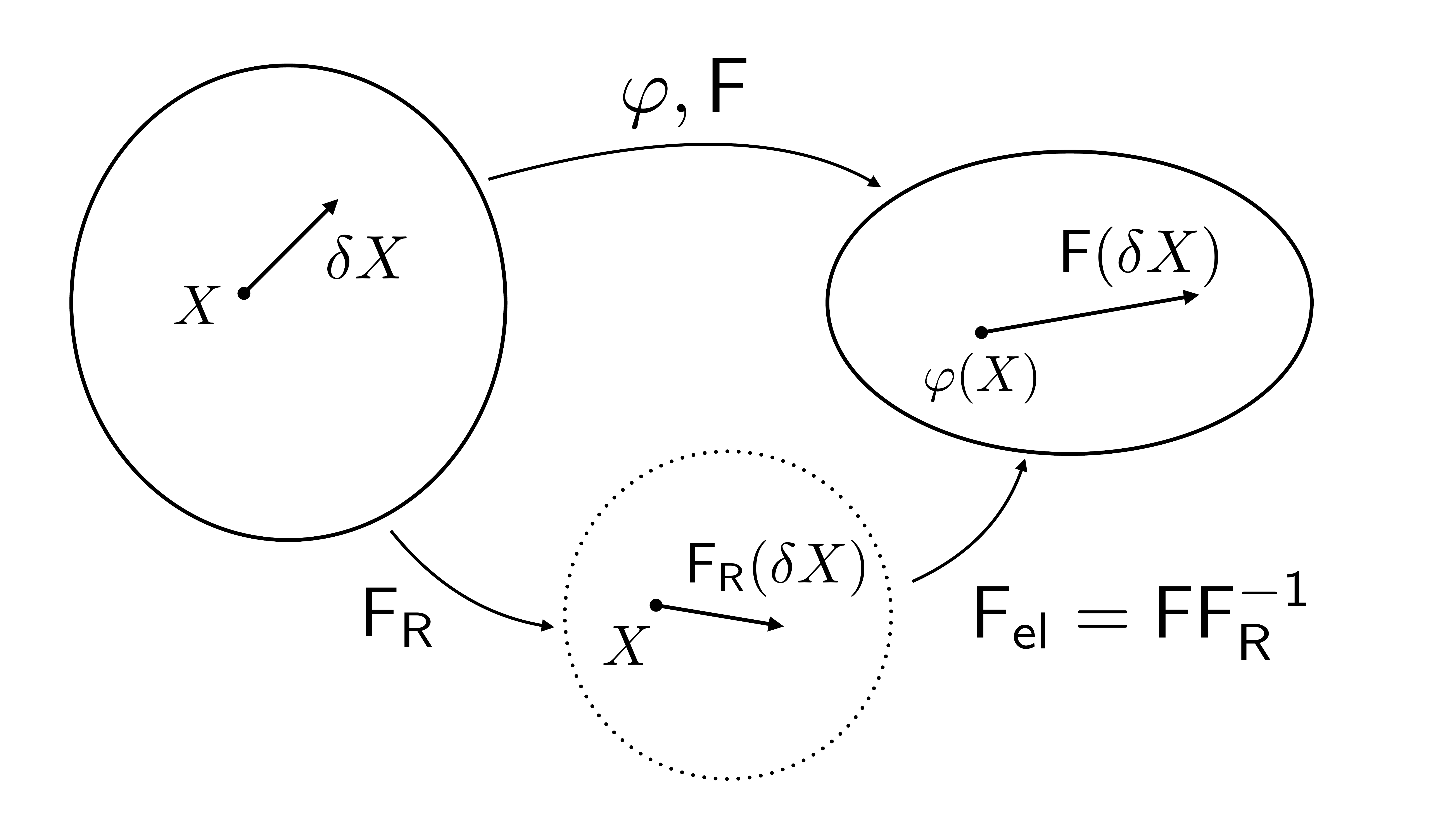}
\caption{Material points $\vc X$ are mapped by the placement $\PL$ into their position in space, while the deformation gradient $\F$ maps tangent vectors $\delta\vc X$ in the material manifold onto spatial vectors. The elastically relaxed state of the material is locally described by $\Fr$, which need not correspond to any global deformation of the material and, in this sense, it maps material vectors onto material vectors. The elastic stresses will then depend upon the mapping $\Fel=\F\Fr^{-1}$ that connects the relaxed material vectors $\Fr(\delta\vc X)$ to the current state of strain given by $\F(\delta\vc X)$.}
\label{fig:F_R}
\end{figure}

An important role in the following construction is played by the deformation gradient $\F$ and related concepts. We recall here some fundamental facts and refer the reader to the book by Gurtin, Fried, and Anand~\cite{Gurtin_2010} for further details. Based on the definition of $\F$ as the gradient with respect to $\vc X$ of the placement $\PL$, we see that $\F$ represents a linear transformation from the tangent space at $\vc X$ to the set of material points into the tangent space at $\PL(\vc X,t)$ to the set of positions. The former contains vectors associated with infinitesimal line elements in the set of $\vc X$ (material line elements), while the latter comprises vectors associated with line elements passing through $\PL(\vc X,t)$ in the ambient space (see Figure~\ref{fig:F_R}, top).

A second basic property of $\F$ is its relation with the velocity gradient. Simply by applying the chain rule to the time derivative of $\F$ one can deduce~\cite{Gurtin_2010} the evolution equation for $\F$ in Eulerian coordinates in the form
\begin{equation}\label{eq:F_evolution}
\Adv{\vc u}\F=(\nabla\vc u)\F.
\end{equation}
Equation \eqref{eq:F_evolution} is an exact kinematic relation between the velocity and the displacement of fluid elements and does not contain any constitutive assumption.
A third fundamental quantity in the description of the deformation is the determinant of $\F$, which corresponds to the coefficient of volumetric dilation from the reference configuration to the current configuration~\cite{Gurtin_2010}. For the isochoric deformations of incompressible materials volumes are always preserved and hence we have $\det\F=1$. This property is preserved by equation \eqref{eq:F_evolution} thanks to the fact that $\tr(\nabla\vc u)=\dvg\vc u=0$.

A continuum model is specified by prescribing constitutive relations between the Cauchy stress $\cauchy$ and the kinematic fields that identify the state of the material, such as $\nabla\vc u$ and $\F$. 
Since we consider incompressible models, we introduce the pressure field $p$ and take $\cauchy=-p\I+\T$, with a pressure term proportional to the identity matrix $\I$ and a treceless extra stress $\T$. 
The latter can be further decomposed in several ways, depending on the effects that one wishes to model. We choose to employ an additive decomposition as sum of a viscous contribution $\Tvi$ plus an elastic one $\Tel$, so that $\cauchy=-p\I+\Tvi+\Tel$.
In what follows, we assume $\Tvi=2\eta \DD$, wherein a viscosity $\eta\geq 0$ multiplies the symmetric part of the velocity gradient $\DD:=(\nabla\vc u+\nabla\vc u^{\tsp})/2$.
At this stage, we pose no restriction on the possibly nonlinear dependence of $\eta$ on other kinematic descriptors.
In later sections we will consider the simplest case of a constant viscosity to highlight the role of the elastic term in determining the model rheology.

To propose a relation for the elastic stress $\Tel$ we need to introduce a tensorial description of the local state of deformation in which no elastic response would be present. Recall that the deformation gradient $\F$ is a linear transformation from the space of material vectors to that of spatial vectors. 
In purely elastic theories, the relaxed state is most often assumed to be realized by a reference material configuration, so that $\F$ maps relaxed (material) line elements into stretched (spatial) line elements. 
We are interested in a more general situation and then we introduce the tensor field $\Fr$, that maps the reference material line elements $\delta\vc X$ into elastically relaxed material line elements. Notice that $\Fr$ need not be the gradient of any deformation: it is a local description of how the material would like to be strained to relax elastic stresses. As a consequence, these stresses will depend upon the mapping $\Fel:=\F\Fr^{-1}$ that connects the relaxed material vectors $\Fr(\delta\vc X)$ to the current state of strain given by $\F(\delta\vc X)$ (see Figure~\ref{fig:F_R}, bottom).

Following this construction of the elastic deformation gradient, we obtain the right and left Cauchy--Green tensors $\Cel:=\Fel^{\tsp}\Fel$ and $\Bel:=\Fel\Fel^{\tsp}$, respectively, sometimes referred to as elastic strain tensors.
The former maps relaxed into relaxed line elements, while the latter maps spatial into spatial line elements. As a consequence, both quantities are objective: $\Bel$ is covariant as $\cauchy$ should be, while $\Cel$ is frame-invariant, as is $\Fr$, given that relaxed line elements are still material ones~\cite{Gurtin_2010}. By introducing the constant elastic modulus $\kappa\geq 0$, we can then postulate
\begin{equation}
\Tel=\kappa\log\Bel=\kappa\log(\F \Fr^{-1}\Fr^{-\tsp}\F^{\tsp}).
\end{equation}
We stress the fact that, by using $\Bel$ in the definition of the stress, we make sure that no elastic response arises when the relative deformation between $\F$ and $\Fr$ is just a rigid rotation.
Moreover, the fact that for incompressible motions we have $\det\F=\det\Fr=1$ implies that $\det\Bel=1$ and so $\tr\Tel=\kappa\tr\log\Bel=0$, as is desirable for a term entering the extra stress.

Note that the actual value of the stress $\Tel$ at any point in time is the outcome of both elastic and inelastic phenomena (the latter being related to plasticity) and it is fully history-dependent. Nevertheless, \revision{at any given time, $\Bel$ is a combination of $\F$ and $\Fr$ that measures precisely the amount of deformation that could be fully recovered. In this sense, $\Tel$ represents instantaneously the elastic part of the extra stress.
The history dependence generated by the plastic evolution of $\Fr$ alongside the viscous stress $\Tvi$ are fundamental in determining the inelastic response. Only the interplay of such terms with the elastic forces can determine the timescales of stress and/or strain recovery observed in different experiments.}

The definition of $\Fel$ given above corresponds to the multiplicative decomposition of the deformation gradient as $\F=\Fel\Fr$, which is part of the standard approach to solid plasticity \cite{Gurtin_2010}. Such a decomposition was introduced by Kr\"oner~\cite{Kroner_1959} and independently by Lee \& Liu~\cite{Lee_1967,Lee_1969} and has since proved to be a key tool also in the description of growth and remodelling~\cite{DiCarlo_2002}, morphogenesis~\cite{Taber_2009}, and active materials~\cite{Nardinocchi_2007}.

In viscoelastic materials, the microscopic arrangement of molecules determines the state at which the system would converge in the absence of applied forces. This state can evolve in time as a result of deformations and stresses. We thus need to postulate a suitable evolution equation for $\Fr$, in keeping with what is customarily done in plasticity theory \cite{Gurtin_2010}. 
Such an equation should imply that, if we keep the material in a static configuration in which some elastic stress is active, then $\Fr$ converges exponentially to $\F$ with a characteristic time-scale $\taur\geq 0$ and the elastic stress relaxes to zero. 
We propose a model based on the following considerations. The rate $(\Adv{\vc u}\Fr)\Fr^{-1}$ is a frame-invariant quantity, because $\Fr$ is frame-invariant, and it belongs to the Lie algebra associated with the group of positive-determinant linear transformations of material line elements into material line elements. This can be seen by considering that any derivative of $\Fr$ takes values in the tangent space to that group at $\Fr$, and the multiplication by $\Fr^{-1}$ maps $\Adv{\vc u}\Fr$ onto an element of the tangent space at the identity, which corresponds to the Lie algebra by definition. Moreover, $(\Adv{\vc u}\Fr)\Fr^{-1}$ must be traceless to preserve the determinant of $\Fr$ during the evolution and, finally, it should vanish whenever the relative deformation between $\F$ and $\Fr$ is a rotation, namely when there is no elastic strain. A quantity that satisfies all of these requirements is $\log\Cel$ and we thus postulate that
\begin{equation}\label{eq:FR_evolution}
\Adv{\vc u}\Fr=\frac{1}{2\taur}(\log \Cel)\Fr.
\end{equation}
The factor of $2$ in front of $\taur$ comes from the fact that $\log\Cel$ contains twice the elastic strain of the material and it is what we need to have the stress relax as $e^{-t/\taur}$ in static experiments.
As is the case for $\eta$, the relaxation time $\taur$ can depend on other quantities that describe the state of the material provided that it remains frame-invariant.

In summary, the evolution equations of the present model are \eqref{eq:F_evolution}, \eqref{eq:FR_evolution}, and the linear momentum balance \eqref{eq:motion} that, upon substitution of our model for the Cauchy stress, becomes
\begin{equation}\label{eq:linear_momentum}
\rho\Adv{\vc u}\vc u=-\nabla p+\dvg(2\eta\vt D+\kappa\log\Bel).
\end{equation}
Boundary conditions must be specified only for the velocity field $\vc u$, while initial conditions are also needed for $\F$ and $\Fr$.

\section{Evolution of the elastic strain and energy balance}\label{sec:energy}

The tensorial measure of elastic strain provided by $\Bel$ satisfies an evolution equation that is determined by those of $\F$ and $\Fr$.
Remembering equations \eqref{eq:F_evolution} and \eqref{eq:FR_evolution}, we get
\[
\vt O=\Adv{\vc u}(\F^{-1}\F)=(\Adv{\vc u}\F^{-1})\F+\F^{-1}(\Adv{\vc u}\F),
\]
from which
\[
\Adv{\vc u}\F^{-1}=-\F^{-1}(\Adv{\vc u}\F)\F^{-1}=-\F^{-1}\nabla\vc u
\]
and $ \Adv{\vc u}\F^{-\tsp}=-\nabla\vc u^\tsp\F^{-\tsp}$, 
and 
\[
\vt O=\Adv{\vc u}(\Fr^{-1}\Fr)=(\Adv{\vc u}\Fr^{-1})\Fr+\Fr^{-1}(\Adv{\vc u}\Fr),
\]
from which
\[
\Adv{\vc u}\Fr^{-1}=-\Fr^{-1}(\Adv{\vc u}\Fr)\Fr^{-1}=-\frac{1}{2\taur}\Fr^{-1}\log (\Cel)
\]
and $\Adv{\vc u}\Fr^{-\tsp}=-\frac{1}{2\taur}\log (\Cel)\Fr^{-\tsp}$.

With these results we can easily compute
\begin{eqnarray}
\Adv{\vc u}\Bel &&=\Adv{\vc u}(\F\Fr^{-1}\Fr^{-\tsp}\F^\tsp)\nonumber\\
&&=\nabla\vc u\Bel+\Bel\nabla\vc u^\tsp\nonumber\\
&&\quad-\frac{1}{\taur}\F\Fr^{-1}(\log\Cel)\Fr^{-\tsp}\F^\tsp.\label{eq:Bel_1s}
\end{eqnarray}
By applying the polar decomposition, we define the orthogonal tensor $\vt R_\mathrm{el}$ via the identity $\F\Fr^{-1}=\Bel^{\frac12}\vt R_\mathrm{el}$.
Moreover, any analytic function $f$ of the tensors $\Cel$ and $\Bel$ is such that $f(\Bel)=\vt R_\mathrm{el}f(\Cel)\vt R_\mathrm{el}^{-1}$, entailing the identity
\begin{equation}\label{eq:Bel_2s}
\F\Fr^{-1}(\log\Cel)\Fr^{-\tsp}\F^\tsp=\Bel^{\frac12}(\log\Bel)\Bel^{\frac12}=\Bel \log\Bel.
\end{equation}
Substituting \eqref{eq:Bel_2s} into \eqref{eq:Bel_1s}, we finally obtain
\begin{equation}\label{eq:Bel_evolution}
\Adv{\vc u}\Bel
=\nabla\vc u\Bel+\Bel\nabla\vc u^\tsp-\frac{1}{\taur}\Bel\log\Bel.
\end{equation}
We stress that the first two terms on the right-hand side of \eqref{eq:Bel_evolution} are of a kinematic nature, since they descend directly from the definition of $\Bel$ and equation \eqref{eq:F_evolution}.
Together with the material derivative they constitute a specific objective rate, which is not postulated but is rather a consequence of basic kinematics.
The last term depends instead on the constitutive choice we made about the dynamics of $\Fr$, intimately linked to the form of the elastic stress as well.
Note that it is possible to derive also en equation for $\log\Bel$, but it would involve more complex and somewhat difficult to interpret nonlinear terms.

The logarithmic nonlinearities that appear in equations \eqref{eq:FR_evolution} and \eqref{eq:Bel_evolution} are structured in the sense that the coefficients at any order of a Taylor expansion are fixed. We find it remarkable that some qualitative features of experiments are recovered by this choice, as we shall see. The choice of a specific nonlinearity can naturally limit the applicability of our model to some classes of materials. Nevertheless, other nonlinear combinations of the logarithmic strain could be used to create further models if the internal interactions of the material required so. 

The form of equation \eqref{eq:Bel_evolution} remains unchanged if $\taur$ depends on other kinematic quantities because the first derivatives of $\Bel$ are expressed in terms of the first derivatives of $\Fr$, proportional to $1/\taur$, and no higher-order derivatives are involved.
Lastly, if $\det\Bel=1$ at the initial time, then it remains such for all times, because
$\tr(\nabla\vc u)=0=\tr(\log\Bel)$.

Equation~\eqref{eq:Bel_evolution} contains in itself less information than the coupled system of \eqref{eq:F_evolution} and \eqref{eq:FR_evolution} but it offers a more direct way to compare our framework with established approaches. In Oldroyd-type models, an evolution for the extra stress or for the elastic stress is postulated. To make a comparison we look for an evolution equation involving $\Tel=\kappa\log\Bel$. By considering the last term of equation \eqref{eq:Bel_evolution}, we see that a good strategy to reconstruct $\Tel$ is to multiply equation \eqref{eq:Bel_evolution} by $\kappa\Bel^{-1}$. On the left-hand side, one might expect to find a logarithmic derivative, nevertheless we have that $\Adv{\vc u}(\log\Bel)$ differs from $\Bel^{-1}\Adv{\vc u}\Bel$ in general, due to the tensorial nature of $\Bel$, and $\Bel$ does not often commute with $\nabla\vc u$. It is only when $\nabla\vc u$ and $\Bel$ are both diagonal on the same \emph{time-independent} basis that we can obtain precisely $\Adv{\vc u}\Tel=2\kappa\vt D-\Tel/\taur$. This shows that in extensional flows we will get results akin to those of the simplest Maxwell model (albeit with a separate viscous contribution). Such a simplification is never true in simple shear or generic flows. A general translation of \eqref{eq:Bel_evolution} would rather be
\begin{equation}\label{eq:Tel_evolution}
\Adv{\vc u}\big(e^{\Tel/\kappa}\big)
=\nabla\vc u\,e^{\Tel/\kappa}+e^{\Tel/\kappa}\nabla\vc u^\tsp-\frac{1}{\kappa\taur} e^{\Tel/\kappa}\Tel,
\end{equation}
which shows a significant structural difference compared to Oldroyd-like models characterized by equations of the form
\[
\Tel+\taur(\Adv{\vc u}\Tel-\nabla\vc u\,\Tel-\Tel \nabla\vc u^\tsp)= f(\vt D,\Tel),
\]
with $f$ a function that takes different forms in different models.
Even considering formulations that employ conformation tensors \cite{Beris_1994} or log-conformation tensors \cite{Fattal_2004}, one would essentially find the same difference. In fact, $\Tel/\kappa$ would be the candidate conformation tensor and we clearly see its exponential appearing in \eqref{eq:Tel_evolution} at places where standard models would put $\Tel/\kappa$.

It is somewhat surprising that the arguments---quite natural from the solid mechanics perspective---that lead to the definition of our framework imply such a dramatic modification when contrasted with Oldroyd's approach.
The reasons to embrace such a change are deeply rooted in the proper mathematical representation of mechanical quantities. Stresses must belong to a linear space, while the dynamics of kinematic quantities such as $\F$, $\Fr$, and $\Bel$, that are quite natural choices to describe the state of the deformation, takes place on tensorial manifolds. Moreover, the evolution of the stress tensor should be a consequence of the evolution of kinematic quantities and constitutive laws.

On the other hand, in the linearized limit of small elastic stresses and strains, we may expect to recover some classical model starting from equation \eqref{eq:Tel_evolution}. Indeed, when $\|\Tel/\kappa\|=\|\log\Bel\|\ll 1$, we can use the approximation $\exp(\Tel/\kappa)\approx \I+\Tel/\kappa$ and substitute it into \eqref{eq:Tel_evolution} to obtain
\begin{eqnarray}
\Adv{\vc u}(\I+\Tel/\kappa)
=&&\nabla\vc u\,(\I+\Tel/\kappa)+(\I+\Tel/\kappa)\nabla\vc u^\tsp\nonumber\\
&&-\frac{1}{\kappa\taur} (\I+\Tel/\kappa)\Tel,\label{eq:linearisation}
\end{eqnarray}
that gives, in the case of constant $\kappa$ and $\taur$, the linearized equation
\begin{equation}\label{eq:UCM}
\Tel+\taur(\Adv{\vc u}\Tel-\nabla\vc u\,\Tel-\Tel\nabla\vc u^\tsp)
=2\kappa\taur\vt D.
\end{equation}
Equation \eqref{eq:UCM} corresponds to the upper-convected Maxwell model~\cite{Bird_1987} with relaxation time $\lambda_1=\taur$ and viscosity $\eta_0=\kappa\taur$. This shows that our model is a proper generalization of Maxwell's model to the case of finite elastic strains. Furthermore, retaining the last term of equation \eqref{eq:linearisation}, which is quadratic in $\Tel$, we find the Giesekus model  \cite{giesekus_1982} with a specific choice of material parameters.

It is finally instructive to compute the energy balance implied by our model.
For a smooth divergence-free velocity field $\vc u$ that satisfies homogeneous boundary conditions in a domain $\Omega$, we can multiply equation \eqref{eq:linear_momentum} by $\vc u$ and integrate by parts to obtain
\begin{eqnarray}
&&\frac{d}{dt}\bigg(
\int_\Omega \frac{\rho}{2}|\vc u|^2\,d\vc x
\bigg)\nonumber\\
&&=\int_\Omega p\dvg\vc u\,d\vc x-\int_\Omega (2\eta\vt D+\kappa\log\Bel):\nabla\vc u\,d\vc x\nonumber\\
&&=-\int_\Omega 2\eta|\vt D|^2\,d\vc x-\int_\Omega \kappa\log\Bel:\vt D\,d\vc x,
\label{eq:der_kinetic_energy}
\end{eqnarray}
where we used the matrix scalar product $\vt A:\vt C:=\tr(\vt A^\tsp\vt C)$, with respect to which symmetric and antisymmetric tensors are orthogonal, and the notation $|\vt A|^2=\vt A:\vt A$. 
Then we can reconstruct the time derivative of the stored elastic energy by multiplying equation \eqref{eq:Bel_evolution} by $(\kappa/2)\Bel^{-1}\log\Bel$, which leads to
\begin{eqnarray}
&&\frac{d}{dt}\bigg(
\int_\Omega\frac{\kappa}{4}|\log\Bel|^2\,d\vc x
\bigg)\label{eq:der_elastic_energy}\\
&&=\int_\Omega \kappa\log\Bel:\vt D\,d\vc x
-\int_\Omega\frac{\kappa}{2\taur}|\log\Bel|^2\,d\vc x.\nonumber
\end{eqnarray}
By taking the sum of \eqref{eq:der_kinetic_energy} and \eqref{eq:der_elastic_energy}, we find
\begin{eqnarray}
&&\frac{d}{dt}\bigg(\int_\Omega \frac{\rho}{2}|\vc u|^2\,d\vc x+
\int_\Omega\frac{\kappa}{4}|\log\Bel|^2\,d\vc x
\bigg)\label{eq:der_total_energy}\\
&&=-\int_\Omega 2\eta|\vt D|^2\,d\vc x
-\int_\Omega\frac{\kappa}{2\taur}|\log\Bel|^2\,d\vc x\leq 0,\nonumber
\end{eqnarray}
showing that the total energy, sum of kinetic and elastic contributions, cannot increase in time and it is lowered via either viscous dissipation, proportional to $|\vt D|^2$, or plastic dissipation, proportional to $|\log\Bel|^2$. 
This proves that the class of models we introduced in the previous section is thermodynamically consistent. It is worth noticing that the cancellation of the exchange term occurring when we sum the two equations stems from the perfect balance between the elastic stress and the plastic evolution.
Finally, we mention some necessary conditions for thermodynamic consistency that may be relevant in generalizing our model. For our elastic stress to limit, and not favor, strain increments, it is necessary to assume $\kappa>0$. Then to ensure that the dissipation is non-negative it is necessary that $\eta$ and $\taur$ remain always non-negative, even in situations where they can vary as functions of other quantities, as we will see in later sections.

\section{Small-amplitude oscillatory flows}\label{sec:small_amplitude}

To investigate the viscoelastic behaviour of the constitutive model introduced in Section~\ref{sec:constitutive} under the simplest assumption of constant $\eta$ and $\taur$, we should first analyze its prediction in small-amplitude oscillatory shear (SAOS) experiments.
With our model, the evolution equation for $\Fr$ in oscillatory shear cannot be easily solved in an analytical way. Nevertheless, we can appeal to the fact that, due to the linearisation implied in the small-amplitude analysis, the results obtained in simple shear are equivalent to those obtained from a small-amplitude oscillatory extensional flow, up to a 45-degree rotation of coordinates. 
In fact, the deformation map in shear, $\PL_\mathrm{sh}$, and extension, $\PL_\mathrm{ex}$, are given by
\[
\PL_\mathrm{sh}(t,x,y)=\begin{pmatrix}
  x+{\gamma(t;\omega, \gamma_0)}y  \\
  y
\end{pmatrix}
\]
and
\[
\PL_\mathrm{ex}(t,x,y)=\begin{pmatrix}
  e^{\frac12\gamma(t;\omega, \gamma_0)} x  \\
  e^{-\frac12\gamma(t;\omega, \gamma_0)} y
\end{pmatrix},
\]
 where the strain is given by $\gamma(t;\omega, \gamma_0) =\gamma_{0} \sin (\omega t)$, with $\gamma_0$ representing the maximum strain amplitude.
The corresponding rate of deformation tensors, connected by a 45-degree rotation, are
\[
\vt D_\mathrm{sh}=
\begin{pmatrix}
0 & \dot\gamma/2\\
\dot\gamma/2 & 0
\end{pmatrix}
\quad\text{and}\quad
\vt D_\mathrm{ex}=
\begin{pmatrix}
\dot\gamma/2 & 0 \\
0 & -\dot\gamma/2
\end{pmatrix},
\]
and a similar relation holds between the linearized Cauchy--Green tensors, that read
\[
\F_\mathrm{sh}\F^\tsp_\mathrm{sh}\approx
\begin{pmatrix}
1 & \gamma\\
\gamma & 1
\end{pmatrix}
\quad\text{and}\quad
\F_\mathrm{ex}\F^\tsp_\mathrm{ex}\approx
\begin{pmatrix}
1+\gamma & 0 \\
0 & 1-\gamma
\end{pmatrix}.
\]

We can then restrict attention to the extensional flow case, setting
\[
 \F(t;\omega,\gamma_0)=\F_\mathrm{ex}=
\begin{pmatrix}
 e^{\gamma(t)/2} & 0  \\
 0 & e^{-\gamma(t)/2}
\end{pmatrix}
\]
and assuming
\[
\Fr(t)=
\begin{pmatrix}
 e^{\gammar(t)/2} & 0  \\
 0 & e^{-\gammar(t)/2}
\end{pmatrix}.
\]
Under these provisions, the evolution equation for $\Fr$ reduces to the scalar evolution equation
\begin{equation}
\frac{\de\gammar}{\de t}=\frac{1}{\taur}(\gamma_{0}\sin(\omega t)-\gammar),
\end{equation}
that, assuming the initial condition $\gammar(0)=0$, leads to
\begin{equation*}
\gammar(t)=\frac{\gamma_{0} \omega\taur  }{1+\omega^{2}\taur^{2} }e^{-t/\taur}+\frac{\gamma_{0} [\sin (\omega t )-\omega\taur   \cos (\omega t )]}{1+\omega^{2}\taur^{2} },
\end{equation*}
which, for sufficiently large times $t\gg \taur$, reduces to
\begin{equation}
\gammar(t)\approx \gamma_{0}\frac{\sin (\omega t )-\omega\taur   \cos (\omega t )}{1+\omega^{2}\taur^{2} }.
\end{equation}

Regarding the elastic strain tensors, we find
\[
\Bel=\Cel=
\begin{pmatrix}
 e^{\gamma-\gammar} & 0  \\
 0 & e^{-(\gamma-\gammar)} 
\end{pmatrix}
\]
and the elastic stress becomes
\[
\Tel(t;\omega,\gamma_0,\taur,\kappa)=\kappa\log\Bel=\kappa(\gamma-\gammar)
\begin{pmatrix}
 1 & 0 \\
 0 & -1 \\
\end{pmatrix},
\]
while the deformation rate tensor is
\[
\DD(t;\omega,\gamma_0)=\frac{\gamma_{0}\omega}{2}\cos(\omega t)\hat{\vt D},\quad\text{with}\quad \hat{\vt D}:=
\begin{pmatrix}
 1 & 0 \\
 0 & -1
\end{pmatrix}.
\]

We can then compute the time-dependent projection
\begin{eqnarray}
&& \zeta(t;\omega,\gamma_0,\taur,\kappa,\eta)  = \frac{(\Tel+\Tvi):\hat{\DD}}{\hat{\DD}:\hat{\DD}}\nonumber\\
&&=\kappa (\gamma-\gammar)+\eta\gamma_0\omega\cos(\omega t)\nonumber\\
&& \stackrel{t\gg\taur}{\approx}\gamma_0\kappa \left[\frac{\omega^2\taur^2}{1+\omega^2\taur^2}\sin(\omega t)+\frac{\omega\taur}{1+\omega^2\taur^2}\cos(\omega t)\right]\nonumber\\
&&\qquad + \gamma_0 \eta \omega  \cos(\omega t)
\end{eqnarray}
which characterizes the material response in this geometry.

The standard material functions for SAOS experiments are $\eta'$, $\eta''$, $G'$, $G''$ defined by
\begin{eqnarray}
&&G'\sin(\omega t)+G''\cos(\omega t)\nonumber\\
&&=\omega\eta'\cos(\omega t)+\omega\eta''\sin(\omega t)=\zeta/\gamma_0.
\end{eqnarray}
For our model we thus find
\begin{eqnarray}
G'(\omega;\taur,\kappa)&&=\omega\eta''(\omega;\taur,\kappa)\nonumber\\
&&=
\omega\frac{\kappa\omega\taur^{2}}{1+\omega^2\taur^2}
,\label{eq:G1}\\
G''(\omega;\taur,\kappa,\eta)&&=\omega\eta'(\omega;\taur,\kappa,\eta)\nonumber\\
&&=\omega\frac{\kappa\taur}{1+\omega^2\taur^2}+\omega\eta.\label{eq:G2}
\end{eqnarray}

\begin{figure*}[t]
\centering
\includegraphics[width=0.9\textwidth]{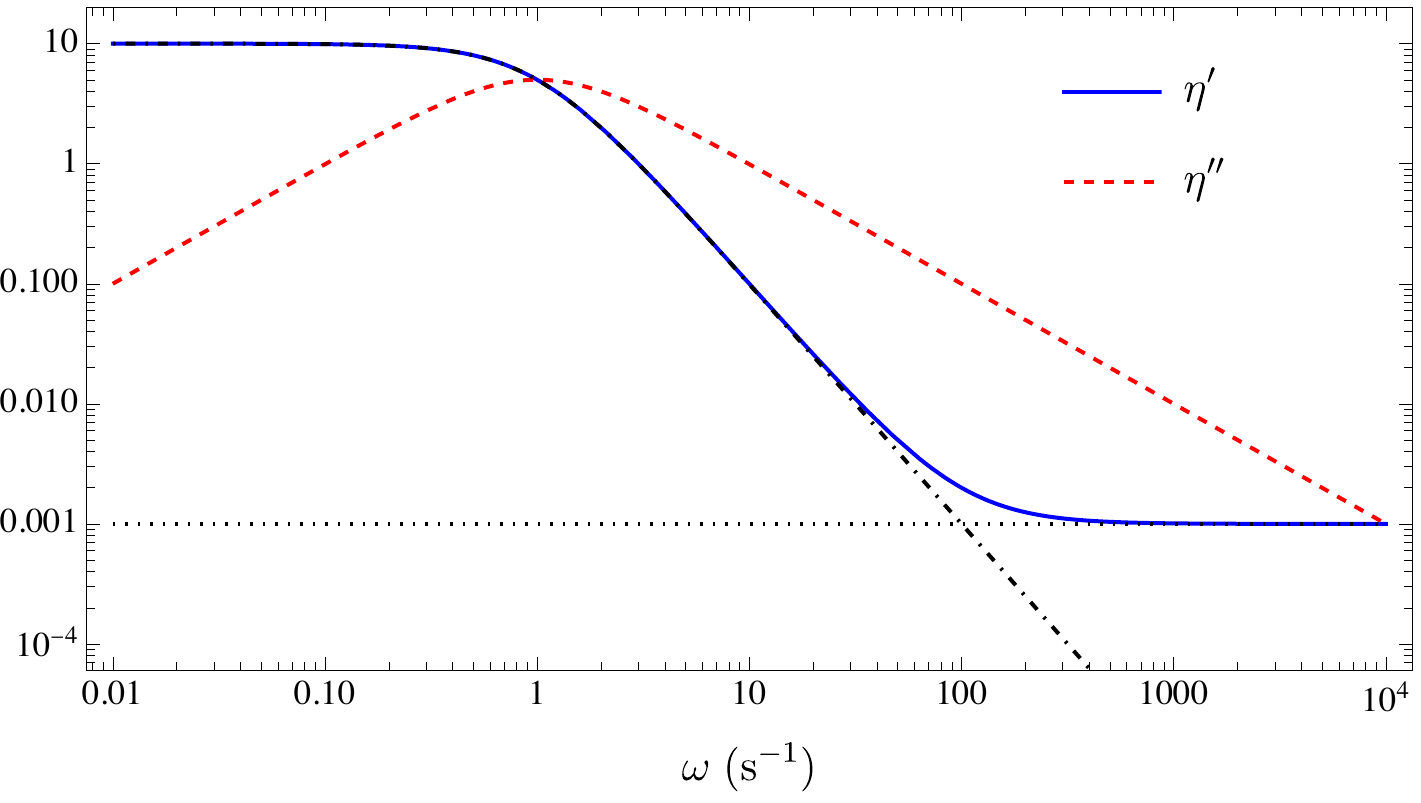}
\caption{\emph{The dependence on the angular frequency $\omega$ of the linear viscoelastic moduli $\eta'$ and $\eta''$ of the present model is typical of systems with a single relaxation time.} The dissipative modulus $\eta'$ is the sum of a plastic contribution (dot-dashed line), identical to the prediction of an upper-convected Maxwell model, and a constant viscous term (dotted line) given by $\eta$. 
The elastic modulus $\eta''$ grows linearly, displays the expected peak around $\omega=1/\taur$, and then decays as $1/\omega$.
In this example, we chose $\taur=1$ s, $\kappa=10$ Pa, and $\eta=10^{-3}$ Pa\,s.
}\label{fig:saos}
\end{figure*}

When $\eta$ is negligible compared to $\kappa\taur$, that is we measure a purely plastic dissipation, the material functions $\eta'$ and $\eta''$ become identical to those predicted by an upper-convected Maxwell model with zero-rate viscosity $\eta_0=\kappa\taur$ (see Figure~\ref{fig:saos}). 
If $\eta$ is smaller than $\kappa\taur$, but not negligible, we obtain a high-frequency plateau in $\eta'$, that does not appear in $\eta''$. 
This is compatible with experimentally observed behaviour that may be fitted by means of Giesekus models (compare, for instance, our Figure~\ref{fig:saos} with Figures 3.4-4 and 3.4-5 of Ref.~\onlinecite{Bird_1987}). In the linear regime, several models collapse on the same curves given by
\begin{gather*}
G'(\omega)=
G\frac{(\lambda\omega)^2}{1+(\lambda\omega)^2},\\
G''(\omega)=
G\frac{\lambda\omega}{1+(\lambda\omega)^2}+\omega\eta_\mathrm{s},
\end{gather*}
where $G$ is the high-frequency storage modulus, $\lambda$ the relaxation time, and $\eta_\mathrm{s}$ the high-frequency viscosity. This is the case for the Johnson--Segalman \cite{Johnson_1977}, single-mode Phan-Thien--Tanner \cite{thien_1977}, and Giesekus \cite{giesekus_1982} models and, manifestly, for the constant-parameter version of our model with the identifications $G=\kappa$, $\lambda=\taur$, and $\eta_\mathrm{s}=\eta$. Notably, those models feature other parameters that do not affect the linear regime, while our model \emph{in the case of constant parameters} is fully determined by the linear viscoelastic spectrum.

The interpretation of the $\omega$-dependence of the dissipative modulus $\eta'$ is the following: when oscillations are slow, $\Fr$ has enough time to evolve and we measure a significant plastic dissipation; when oscillations are fast, $\Fr$ remains effectively fixed and the material behaves like a viscoelastic solid, where the measured dissipation is purely viscous.
Regarding the elastic response, it is easier to interpret the behaviour of $G'=\omega\eta''$: if the relaxation time $\taur$ is much shorter than $1/\omega$, then the elastic response is negligible and $G'\propto \omega^2\taur^2$; when $\taur\gg 1/\omega$, then $\Fr$ remains effectively fixed and $G'\approx \kappa$ is independent of $\omega$.
Hence, with a sufficiently broad range of frequency, oscillatory flow experiments would allow to measure all of the material parameters.

\section{Stress growth after inception of steady flows}\label{sec:stress_growth}

In this section, we analyze the evolution of the stress when a steady flow is suddenly imposed starting from a stress-free static condition. The constant strain rate is $\gd:=\sqrt{2\DD:\DD}$. The components of the extra stress are represented by the following time-dependent material coefficients
\begin{equation}\label{eq:mat_coeff+}
\begin{aligned}
\etaeff^+(t;\taur,\kappa,\eta,\gd):=&\frac{1}{\gd}\frac{\T:\hat{\DD}}{\hat{\DD}:\hat{\DD}},\\ 
\psieff^+(t;\taur,\kappa,\eta,\gd):=&\frac{2}{\gd^2}\frac{\T:\hat{\G}}{\hat{\G}:\hat{\G}},
\end{aligned}
\end{equation}
where, for planar flows, we set
\[
\hat{\DD}:=\frac{2}{\gd}\DD,
\;
\hat{\G}:=\frac12(\vt A\hat{\DD}-\hat{\DD}\vt A),\text{ and }\vt A:=
\begin{pmatrix}
0 & 1 \\
-1 & 0
\end{pmatrix}.
\]
Following Giusteri \& Seto~\cite{Giusteri_2018}, we employ a definition of the material coefficients that is independent of the flow type, to be able to directly compare the results obtained in extensional and simple shear flows.

The definitions given in \eqref{eq:mat_coeff+} coincide with the standard material functions (viscosity and first normal stress coefficient) in simple shear flows, but can be used to analyze any steady flow.
Steady-state material functions will be determined as the long-time limit of $\etaeff^+$ and $\psieff^+$.
We will find analytical expressions in planar and uniaxial extensional flows and then numerically study the case of simple shear.

\begin{figure*}[t]
\centering
\includegraphics[width=0.9\textwidth]{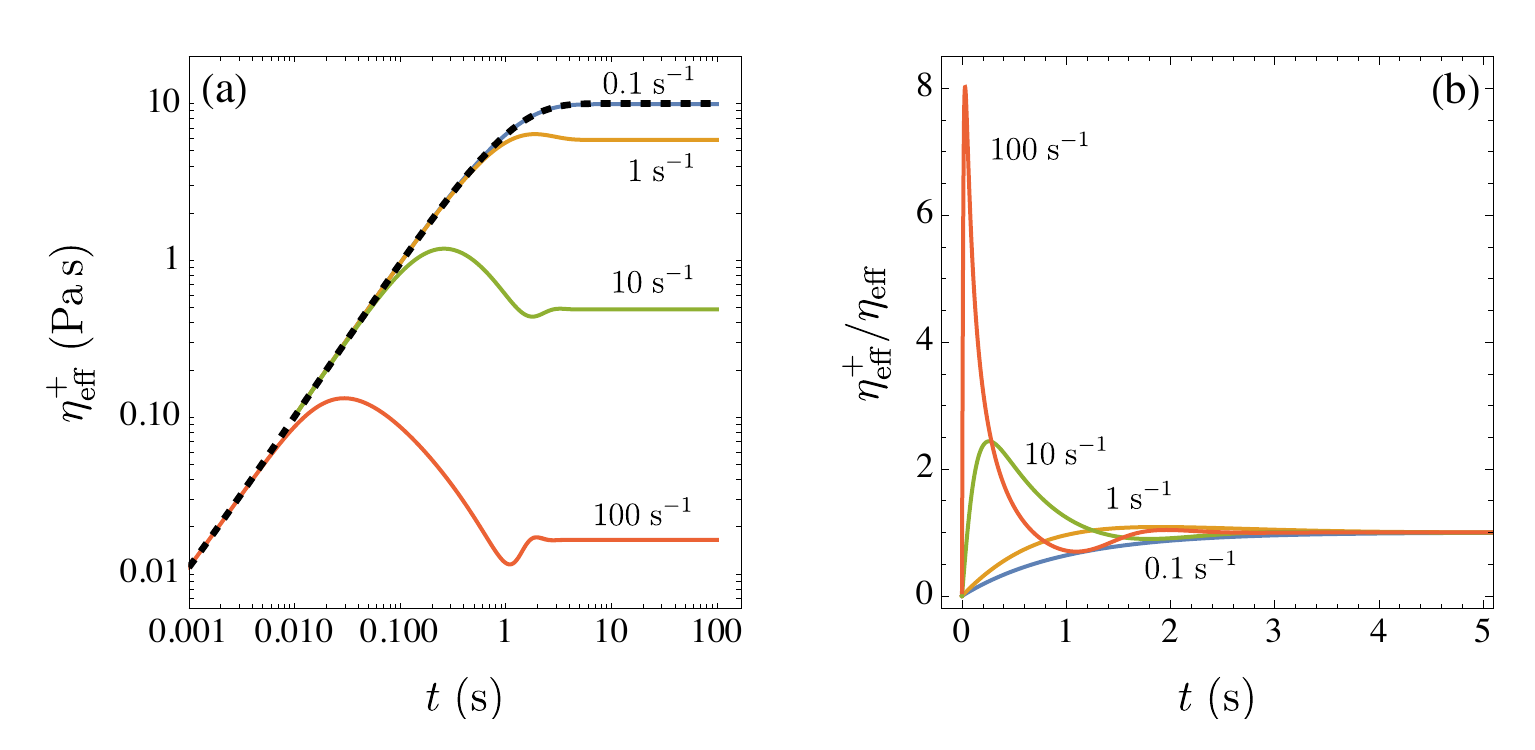}
\caption{\emph{Stress overshoot after inception of steady shear is well captured together with a rate-dependent behaviour.}
Looking at the time-evolution of $\etaeff^+$ for different values of the shear rate $\gd$ (curve labels), we see that in the low-rate regime, with $\gd\ll 1/\taur$, the rate-dependence becomes negligible and $\etaeff^+$ converges to the extensional flow result (black dashed line in panel (a)). At high shear rates, $\gd\gtrsim 1/\taur$, the growth of $\etaeff^+$ is hindered, it reaches a maximum at a strain $\gd t\approx 3$, and then decays toward a rate-dependent asymptotic value, with possibly a few oscillations. 
This behaviour is even clearer by looking at the data normalized, in panel (b), by the steady value $\etaeff$. The values of material parameters are $\taur=1$ s, $\kappa=10$ Pa, and $\eta=10^{-3}$ Pa\,s.}\label{fig:etaplus}
\end{figure*}

\subsection{Planar extensional flow}

In this case, the deformation map is
\[
\PL(t, x,y)=\begin{pmatrix}
  e^{\frac12\gd t} x  \\
  e^{-\frac12\gd t} y
\end{pmatrix},
\] 
the deformation gradient
\[
\F=
\begin{pmatrix}
 e^{\frac12\gd t} & 0  \\
 0 & e^{-\frac12\gd t} 
\end{pmatrix},
\]
and the relaxed state
\[
\Fr=
\begin{pmatrix}
 e^{\frac12\gammar(t)} & 0 \\
 0 &e^{-\frac12\gammar(t)} 
\end{pmatrix}.
\]

The evolution equation for $\Fr$ reduces to the scalar equation
\begin{equation}\label{eq:gammaR}
\frac{\de\gammar}{\de t}=\frac{1}{\taur}(\gd t-\gammar),
\end{equation}
that, with the initial condition $\gammar(0)=0$, leads to
\begin{equation}\label{eq:sol_gammaR}
    \gammar(t)=\gd\left( t-\taur + \taur e^{-t/\taur}\right).
\end{equation}
From this, we easily obtain
\begin{equation}
\log\Bel=\taur  \gd \left(1-e^{-t/\taur }\right)\begin{pmatrix}
 1 & 0 \\
 0 & -1
\end{pmatrix}.
\end{equation}
Recalling that in this case the deformation rate tensor is
\[
\DD= \frac{\dot{\gamma}}{2} \begin{pmatrix}
 1 &0\\
  0 &  -1  \\
\end{pmatrix} ,
\]
we find
\begin{equation}\label{eq:mat_coeff+_extensional}
\begin{aligned}
\etaeff^+(t;\taur,\kappa,\eta,\gd)&=
\kappa\taur \left(1-e^{-t/\taur }\right)+\eta,\\ 
\psieff^+(t;\taur,\kappa,\eta,\gd)&=0.
\end{aligned}
\end{equation}

We stress that $\etaeff^+$ turns out to be independent of the shear rate $\gd$, and so is the steady-state apparent viscosity
\[
\etaeff(\taur,\kappa,\eta):=\lim_{t\to+\infty}\etaeff^+(t;\taur,\kappa,\eta)=\kappa\taur+\eta,
\]
which is also identical to the zero-frequency dissipative modulus $\eta'(0;\taur,\kappa,\eta)$.
The normal stress coefficient $\psieff^+$ is rate-independent in a trivial way: it is identically zero in extensional flows due to symmetry reasons.

\subsection{Uniaxial extensional flow}

In uniaxial extensional flow the deformation map  $\PL$ is given by \[
\PL(t,x,y,z)=(e^{-\frac{\gd t}{4} }x,e^{-\frac{\gd t}{4} }y,e^{\frac{\gd t}{2} }z)^\tsp,
\]
and the deformation gradient $\F$ and the relaxed state $\Fr$ are given by
\begin{gather*}
\F=
\begin{pmatrix}
  e^{-\gd t/4 } & 0 & 0 \\
 0 & e^{-\gd t/4 } & 0 \\
 0 & 0 & e^{\gd t/2 } \\
\end{pmatrix},\\
\Fr=
\begin{pmatrix}
 e^{-\gammar/4 } & 0 & 0 \\
 0 & e^{-\gammar/4 } & 0 \\
 0 & 0 & e^{\gammar/2 } \\
\end{pmatrix}.
\end{gather*}
The evolution equation for $\Fr$ reduces again to \eqref{eq:gammaR} with solution \eqref{eq:sol_gammaR}.
Since we obtain
\[
\log\Bel=\gd\taur\left(1-e^{-t\taur }\right) 
\begin{pmatrix}
 -1/2 & 0 & 0 \\
 0 & -1/2 & 0 \\
 0 & 0 & 1
\end{pmatrix}
\]
and
\[
\DD=\frac{\gd}{2} 
\begin{pmatrix}
 -1/2 & 0 & 0 \\
 0 & -1/2 & 0 \\
 0 & 0 & 1
\end{pmatrix},
\]
we eventually arrive at a result which is identical to the one obtained for planar extensional flows.

\begin{figure*}[t]
\centering
\includegraphics[width=0.9\textwidth]{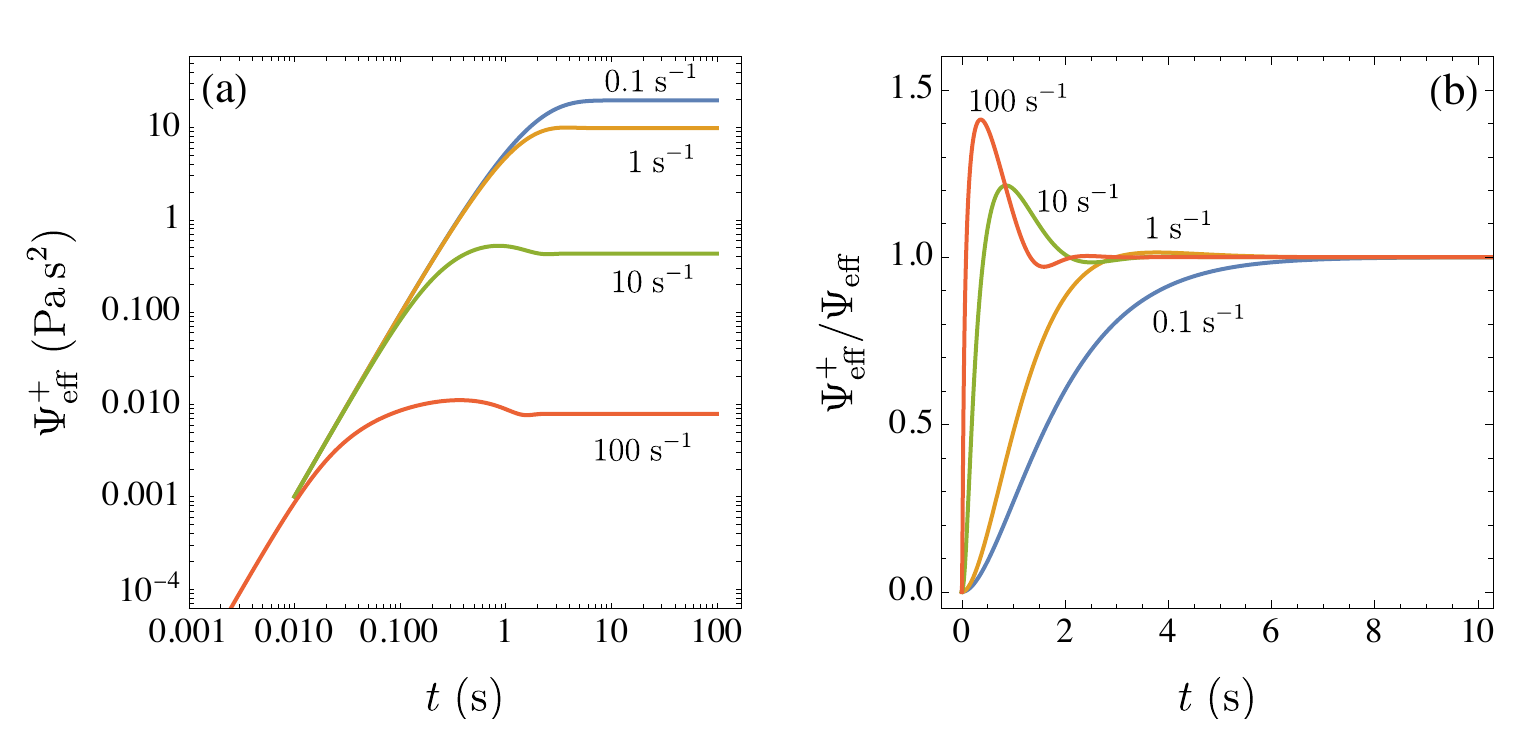}
\caption{\emph{The first normal stress coefficient $\psieff^+$ is nonzero in simple shear and its growth after inception of steady shear is rate-dependent.} At low rates ($\gd<<1/\taur$) the rate-dependence disappears and the growth is monotonic, while for $\gd\gtrsim 1/\taur$ we observe a short-time peak followed by a decay toward the rate-dependent steady value $\psieff$ (panel (a)). 
This behaviour is emphasized in the data normalized by the steady value $\psieff$, shown in panel (b). The values of material parameters are $\taur=1$ s, $\kappa=10$ Pa, and $\eta=10^{-3}$ Pa\,s.}\label{fig:psiplus}
\end{figure*}

\subsection{Simple shear}

For simple shear flows, the deformation map $\PL$ and the deformation gradient $\F$ are given by
\[
\PL(t, x,y,\gd)=\begin{pmatrix}
x + y\gd t \\
y
\end{pmatrix}
\quad\text{and}\quad
\F=
\begin{pmatrix}
1 & \gd t  \\
 0 & 1 
\end{pmatrix},
\]
but there is no reason to assume a specific shape for the relaxed state $\Fr$, which must be determined by solving the evolution equation \eqref{eq:FR_evolution}.
This is a fully tensorial evolution equation that cannot be reduced to a scalar one. We then solve it numerically. Since the flow is homogeneous, $\Fr$ is independent of the space variables and the convective derivative reduces to an ordinary time derivative. We then apply an explicit Euler method for the time integration. This combines the finite-difference approximation of the time derivative with the evaluation of the nonlinear right-hand side of equation \eqref{eq:FR_evolution} at the initial time for each interval of the temporal discretization. Once a time step $\varDelta t$ is chosen, the temporal mesh is given by the times $t_n=n\varDelta t$, for $n\in\mathbb N$, and the value of $\Fr$ at $t_{n+1}=t_n+\varDelta t$ is given by 
\[
\Fr(t_{n+1})=\Fr(t_n)+\frac{\varDelta t}{2\taur}[\log\Cel(t_n)]\Fr(t_n),
\]
with $\Cel(t_n)=\Fr(t_n)^{-\tsp}\F^\tsp(t_n)\F(t_n)\Fr(t_n)^{-1}$. The integration of such an equation is very fast and we made sure to choose a time step sufficiently small to guarantee the reliability of the computed solution.

A very important consequence of the structure of equation \eqref{eq:FR_evolution} is that the eigenvectors of $\Bel$ rotate over time and are different from those of $\DD$. This gives rise to two important effects: the first normal stress coefficient is no longer zero and we observe a dependence of the material functions $\etaeff^+$ and $\psieff^+$ on the shear rate $\gd$.
In the low-rate regime, with $\gd\ll 1/\taur$, the rate-dependence becomes negligible and $\etaeff^+$ converges to the extensional flow result \eqref{eq:mat_coeff+_extensional}, which is also coincident with the prediction, for simple shear, of an upper-convected Maxwell model (though this would give a different result in extension).
When, on the other hand, $\gd\gtrsim 1/\taur$, the measured shear stress stems from a competition between the speed of the rotation of stress eigenvectors and that of the plastic relaxation.
Essentially, the growth of $\etaeff^+$ is hindered, it reaches a maximum at a strain $\gd t\approx 3$, and then decays toward a rate-dependent asymptotic value, with possibly a few oscillations (see Figure~\ref{fig:etaplus}).
A similar behaviour is followed by the normal stress coefficient $\psieff^+$ that shows a rate-independent growth for $\gd\ll 1/\taur$ and a rate-dependent asymptotic value for $\gd\gtrsim 1/\taur$ (see Figure~\ref{fig:psiplus}).
Our findings are qualitatively compatible with experimental data as can be seen by comparing them with Figures~3.4-7, 3.4-8, 3.4-9, and 3.4-10 of Ref.~\onlinecite{Bird_1987}. 

\begin{figure*}[t]
\centering
\includegraphics[width=0.9\textwidth]{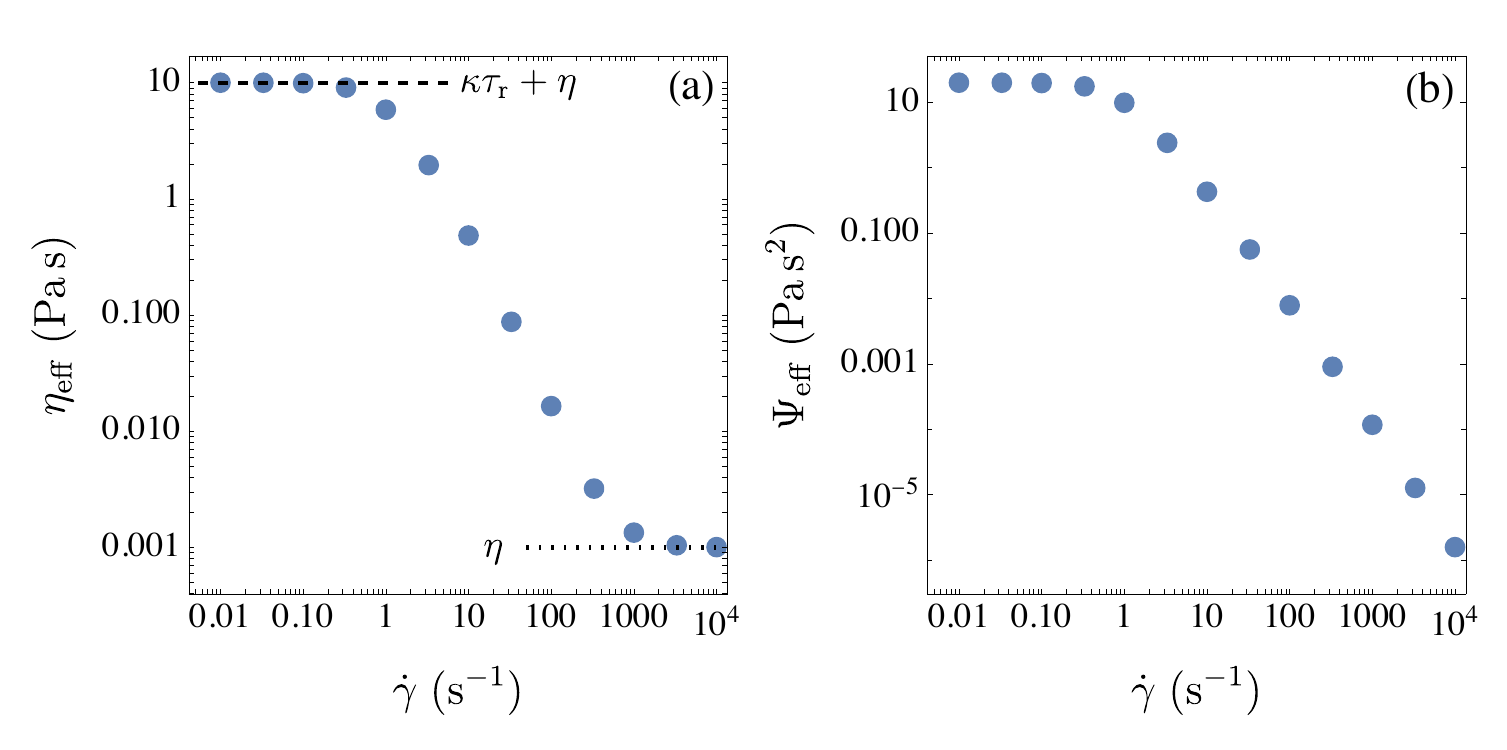}
\caption{\emph{The rate dependence in simple shear of the steady-state material functions $\etaeff$ and $\psieff$ displays a shear-thinning behaviour common to many viscoelastic fluids.} The effective viscosity, in panel (a), interpolates between the zero-rate value, given by $\kappa\taur+\eta$ (dashed line), and $\eta$ (dotted line), while the effective normal stress coefficient, in panel (b), simply decreases to zero.
The values of material parameters are $\taur=1$ s, $\kappa=10$ Pa, and $\eta=10^{-3}$ Pa\,s.
}\label{fig:steadystate}
\end{figure*}

The steady state material functions $\etaeff$ and $\psieff$ display, in simple shear, a rate-dependent behaviour of shear-thinning type (Figure~\ref{fig:steadystate}). The effective viscosity decreases, for $\gd\gtrsim 1/\taur$, from the low-rate value given by $\kappa\taur+\eta$ to the asymptotic value set by $\eta$. We thus see that plastic dissipation dominates the low-rate material response. The effective normal stress coefficient, which is not influenced by the viscosity parameter $\eta$, simply decreases from its zero-rate value and vanishes asymptotically.
Overall, the simple assumptions present in our model with constant parameters can reproduce qualitative features common to several viscoelastic fluids while providing an understanding of the underlying competition between elasticity, relaxation, and flow geometry that cannot be achieved by means of empirical data-fitting laws.

In this section, we treated simple shear as a two-dimensional flow. If, on the other hand, we view it as a planar but three-dimensional flow, a second normal stress coefficient becomes relevant for the characterisation of the material. As long as the deformation remains planar, the present model assigns a vanishing stress in the third spatial direction. This implies that the second normal stress coefficient coincides always with $-\psieff/2$. 
The sign of this quantity is consistent with most measurements in the context of viscoelastic fluids, but the reported magnitude is typically significantly lower~\cite{Maklad_2021}. We believe that further consideration of how to capture experimental observations of second normal stress coefficients will be an important direction for future investigations.

\section{Stress relaxation after a sudden deformation}\label{sec:stress_relax}

A fundamental parameter of our model is the relaxation time $\taur$ and it can be directly related to stress relaxation experiments. We consider a material held in a static configuration after a very rapid homogeneous deformation. In the case of an extensional deformation, we can compute the time-dependent elastic stress by taking
\[
\F=
\begin{pmatrix}
 e^{\frac12\gamma_0} & 0  \\
 0 & e^{-\frac12\gamma_0} 
\end{pmatrix},\qquad \Fr=
\begin{pmatrix}
 e^{\frac12\gammar(t)} & 0 \\
 0 &e^{-\frac12\gammar(t)} 
\end{pmatrix}.
\]
with $\gamma_0$ constant. The evolution equation for $\Fr$ easily leads to $\gammar(t)=\gamma_0(1-e^{-t/\taur})$ and
\begin{eqnarray}
\Tel&&=\kappa\log\Bel=(\gamma_0-\gammar)\begin{pmatrix}
 1 & 0 \\
 0 & -1
\end{pmatrix}\nonumber\\
&&=\kappa\gamma_0e^{-t/\taur }\begin{pmatrix}
 1 & 0 \\
 0 & -1
\end{pmatrix}.
\end{eqnarray}

From this, we clearly see that the stress decays exponentially with rate $1/\taur$. By numerical integration of the evolution equation for $\Fr$, we obtain the same decay also in the case of a simple shear deformation. Nevertheless, in simple shear the prefactor can depend nonlinearly on $\gamma_0$ for large initial strain.
At this point, we have available several sets of experiments with which we can measure the values of the material parameters $\taur$, $\kappa$, and $\eta$.

\begin{figure*}[t]
\centering
\includegraphics[width=0.9\textwidth]{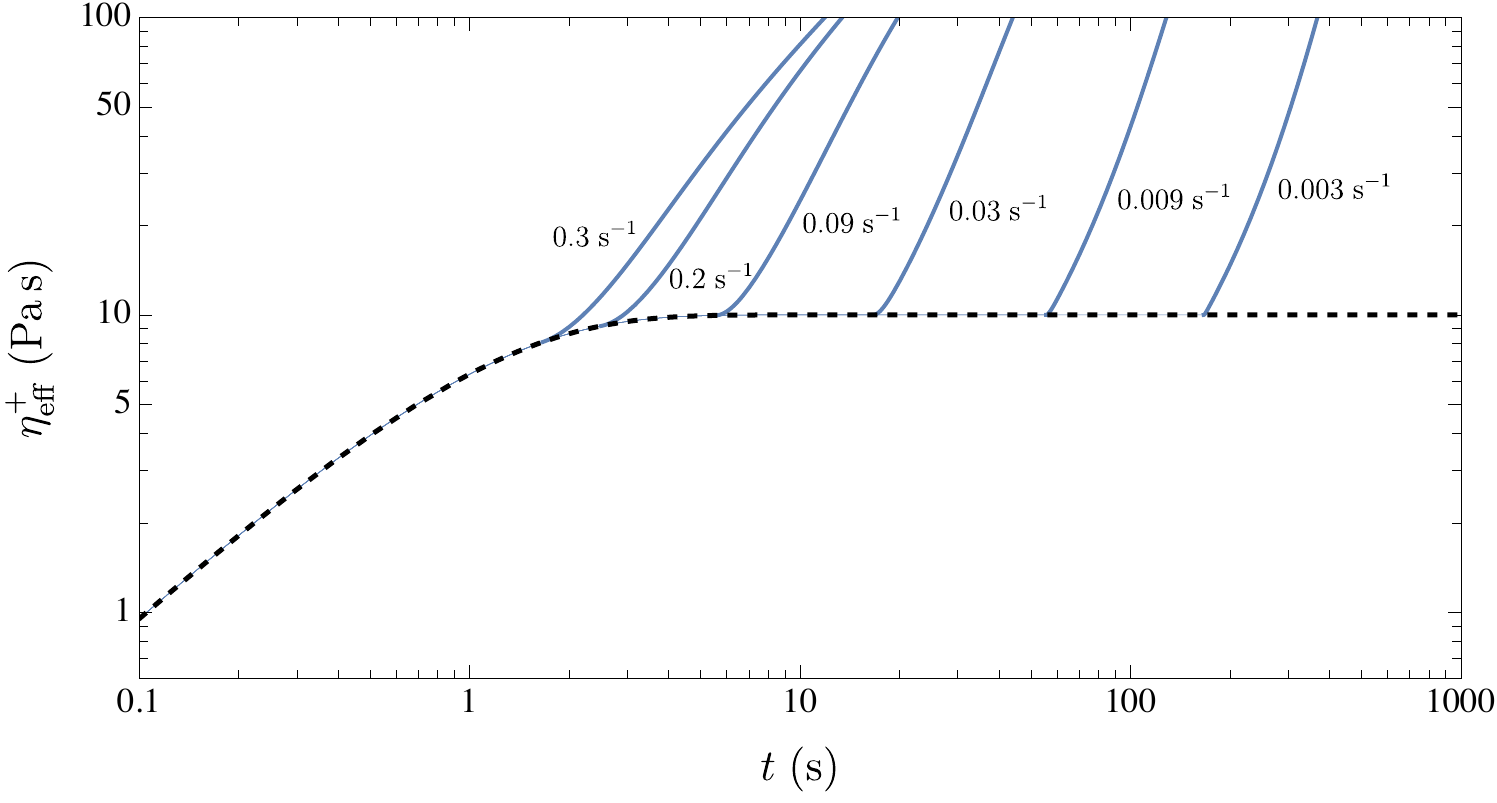}
\caption{\emph{With a strain-dependent relaxation time we can capture polymeric jamming.} The growth of the effective viscosity $\etaeff^+$ in uniaxial extension features a rate-independent segment (black dashed line) followed until the rate-dependent time $\gammaj/\gd$, after which we observe a fast increase (solid lines) due to the rapidly growing relaxation time. Values of $\gd$ label the different data curves.
The values of model parameters for this example are $\taur^0=1$ s, $\gammaj=0.5$, $\alpha=4$, $\kappa=10$ Pa, and $\eta=10^{-3}$ Pa\,s.
}\label{fig:etauniax}
\end{figure*}

\section{Polymeric jamming in uniaxial extension}\label{sec:jamming}

The results of the previous sections show that part of the flow-type dependence of the material response, namely the shear-thinning behaviour versus rate-independence in extensional flows, is not related to microscopic phenomena but rather to the rotation of the principal strains in simple shear (which in this context is not that simple, after all). Nevertheless, there are further differences, such as extensional thickening or the impossibility of reaching a steady state, that cannot be captured with the simplest constant-parameter model.

Here we propose a mechanism and a model that can explain some features of extensional rheology, focusing on uniaxial extension for definiteness. We argue that the typical experimental realization of uniaxial extension can lead, for some fluids, to the phenomenon of polymeric jamming. That is, molecular chains that are mostly elongated in one direction and, due to the confinement in filament stretching experiments, compressed in the orthogonal plane become progressively unable to relax.
We do not expect a similar phenomenon in simple shear flows where confinement does not change over time.
Note that this type of effect is not necessarily related to the bulk rheology of the material, but rather to its interaction with the experimental setup, meaning that it may not appear in large-scale extensional flows.
Certainly, the ultimate explanation of this significant strain-hardening effect requires an investigation of the  microscopic physics that goes far beyond the scope of our study. Here we follow a phenomenological approach to illustrate the possibility of capturing such effects within our modelling framework.

A very rough model for polymeric jamming can be set up by letting the relaxation time $\taur$ depend on a measure of the total strain such as $\gamma(t)=\gd t$, which is obviously related to the filament shrinking.
In particular, we fix a threshold $\gammaj>0$ and we assume that $\taur=\taur^0$, with $\taur^0$ constant and finite, as long as $\gamma(t)<\gammaj$, whereas $\taur=\taur^0e^{\alpha (\gamma(t)-\gammaj)}$ if $\gamma(t)\geq\gammaj$. Note that the energy balance \eqref{eq:der_total_energy} still applies to this model. 
The exponential growth of $\taur$ is, at this stage, an arbitrary modelling choice to have the material approach a solid one ($\taur=+\infty$) rather fast. 
It is possible to solve equation~\eqref{eq:gammaR} analytically in the region with growing $\taur$, but the expression of the solution is not quite informative, and numerical integration is equally effective.
The parameter $\alpha>0$ determines the slope of the effective viscosity $\etaeff^+$ beyond the jamming strain.
Indeed, with this model we predict a rate-independent behaviour in the initial (rate-dependent) interval $[0,\gammaj/\gd]$, with $\etaeff^+(t)=\kappa\taur^0(1-e^{-t/\taur^0})+\eta$, followed by a fast increase, with a slope that looks approximately the same in log-scale for $\gd$ sufficiently small, and no steady state is attained. 
By choosing $\alpha=4$, we obtain a picture, in Figure~\ref{fig:etauniax}, that is in striking qualitative agreement with Series PSII in Figure 3.5-2 of Ref.~\onlinecite{Bird_1987}.
This provides a strong indication that our model can capture the mechanism behind the experimental observation. 

\section{Developing models from experimental data}\label{sec:developing}

As shown above, to reproduce experimental results we generally need to go beyond the constant-parameter model. Especially if we wish to achieve a quantitative description of real fluids, we cannot ignore the presence of multiple relaxation mechanisms.
From the microscopic perspective, we can argue that an increase of the average kinetic energy of the polymeric molecules can activate different relaxation pathways. Neglecting temperature effects, we can include this fact in our macroscopic description by letting the effective relaxation time $\taur$ depend on the local strain rate.

Regarding material functions in the steady-state regime, we can start from the equation solved by $\Bel$, namely
\begin{equation}
\log\Bel=\Bel^{-1}(\taur\nabla\vc u)\Bel+(\taur\nabla\vc u)^\tsp,
\end{equation}
to see that the dependence of the elastic stress on the shear rate $\gd$ is only through the dimensionless Weissenberg number $\Wi=\gd\taur$, even when $\taur$ depends on $\gd$.
From the practical point of view, this allows us to reconstruct the rheological curves for different models of $\taur$ from those obtained with constant unit values of $\taur$ and $\kappa$.

In small-amplitude oscillatory flows, from the solution given in Section~\ref{sec:small_amplitude}, we see that the storage and loss moduli, $G'$ and $G''$, depend on the frequency $\omega$ through the dimensionless Deborah number $\De=\omega\taur$ as long as $\taur$ is independent of time. Again, we can exploit this fact and consider models in which $\taur$ depends on $\omega$. The rationale for this is that the maximum strain rate is proportional to $\omega$ and so the average kinetic energy available to activate relaxation pathways depends on the angular frequency as well. With this in mind, we can discuss further comparisons with experimental data.

\begin{figure*}[t]
\centering
\includegraphics[width=0.9\textwidth]{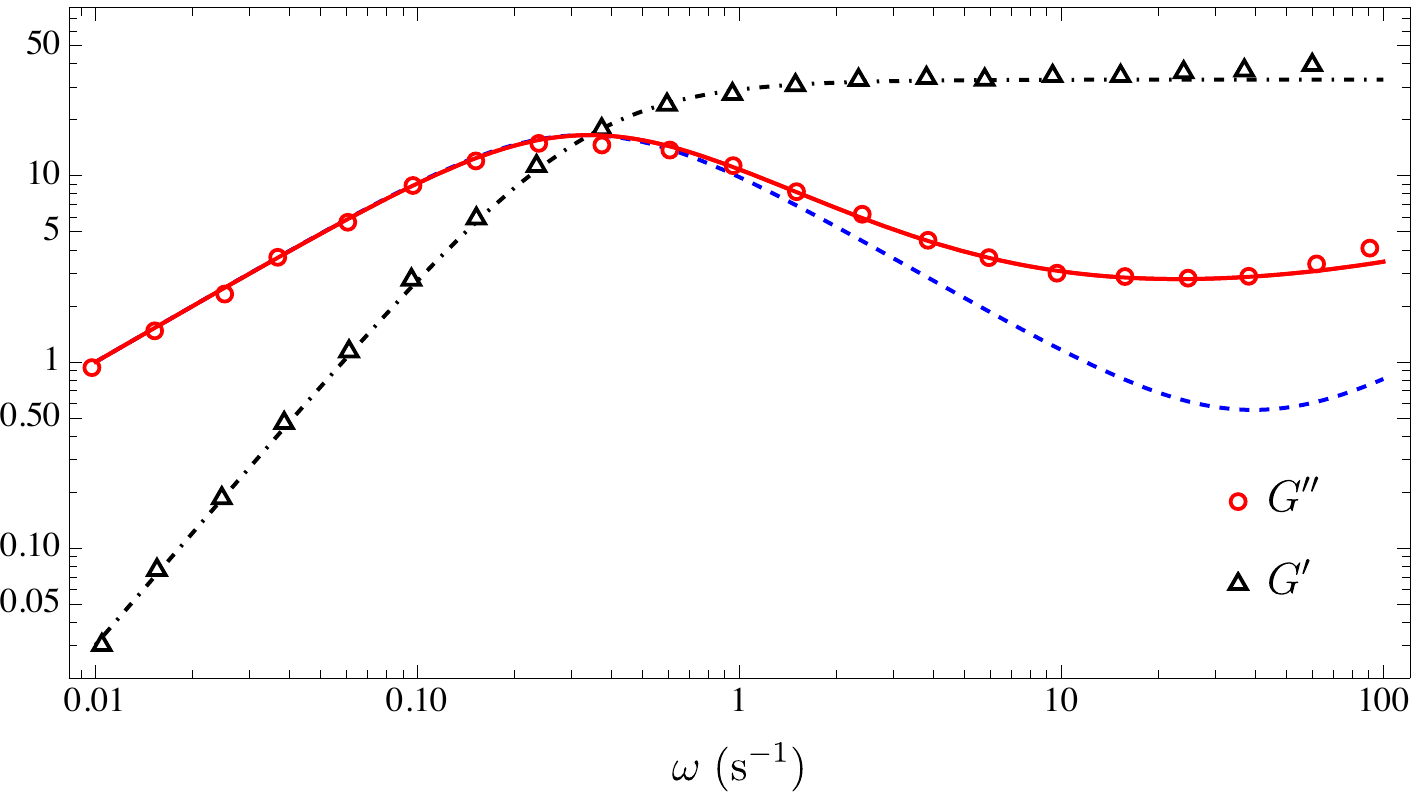}
\caption{\emph{The dependence on the angular frequency $\omega$ of the linear viscoelastic moduli $G'$ and $G''$ for a WLM solution~\cite{Haward_2012} can be well reproduced with an $\omega$-dependent relaxation time.} The solid red line shows the $G''$ curve obtained by assuming relation \eqref{eq:taur_down} with $\taur^0=3$ s, $\alpha=0.165$ s, and $n=1.14$, and setting $\kappa=33$ Pa and $\eta=0.007$ Pa\,s.
The result with a constant $\taur=3$ s is clearly worse and it is the same that one would obtain with several other models~\cite{Varchanis_2022}. The effect on the storage modulus $G'$ of the varying $\taur$ is, in this case, hardly noticeable (dot-dashed line).
}\label{fig:saosWLM}
\end{figure*}

We consider measurements of rheological properties of a wormlike micellar solution (WLM) reported by Haward \& McKinley \cite{Haward_2012} and recently employed by Varchanis \emph{et al.}~\cite{Varchanis_2022} to evaluate the performance of various constitutive models. Regarding the linear viscoelastic properties, constant-parameter models can only capture the data in a partial way and we recover the same results of the Giesekus model, on which also the more sophisticated Vasquez--Cook--McKinley model \cite{Vasquez_2007} collapses (see Figure~\ref{fig:saosWLM}, blue dashed line). If, following the discussion above, we let the relaxation time $\taur$ depend on the frequency according to the phenomenological law
\begin{equation}\label{eq:taur_down}
\taur(\omega)=\frac{\taur^0}{1+(\alpha\omega)^n}
\end{equation}
with appropriate choices of the parameters $\taur^0$, $\alpha$, and $n$, we achieve a much better agreement with experimental data (Figure~\ref{fig:saosWLM}, red solid line).

\begin{figure*}[t]
\centering
\includegraphics[width=0.9\textwidth]{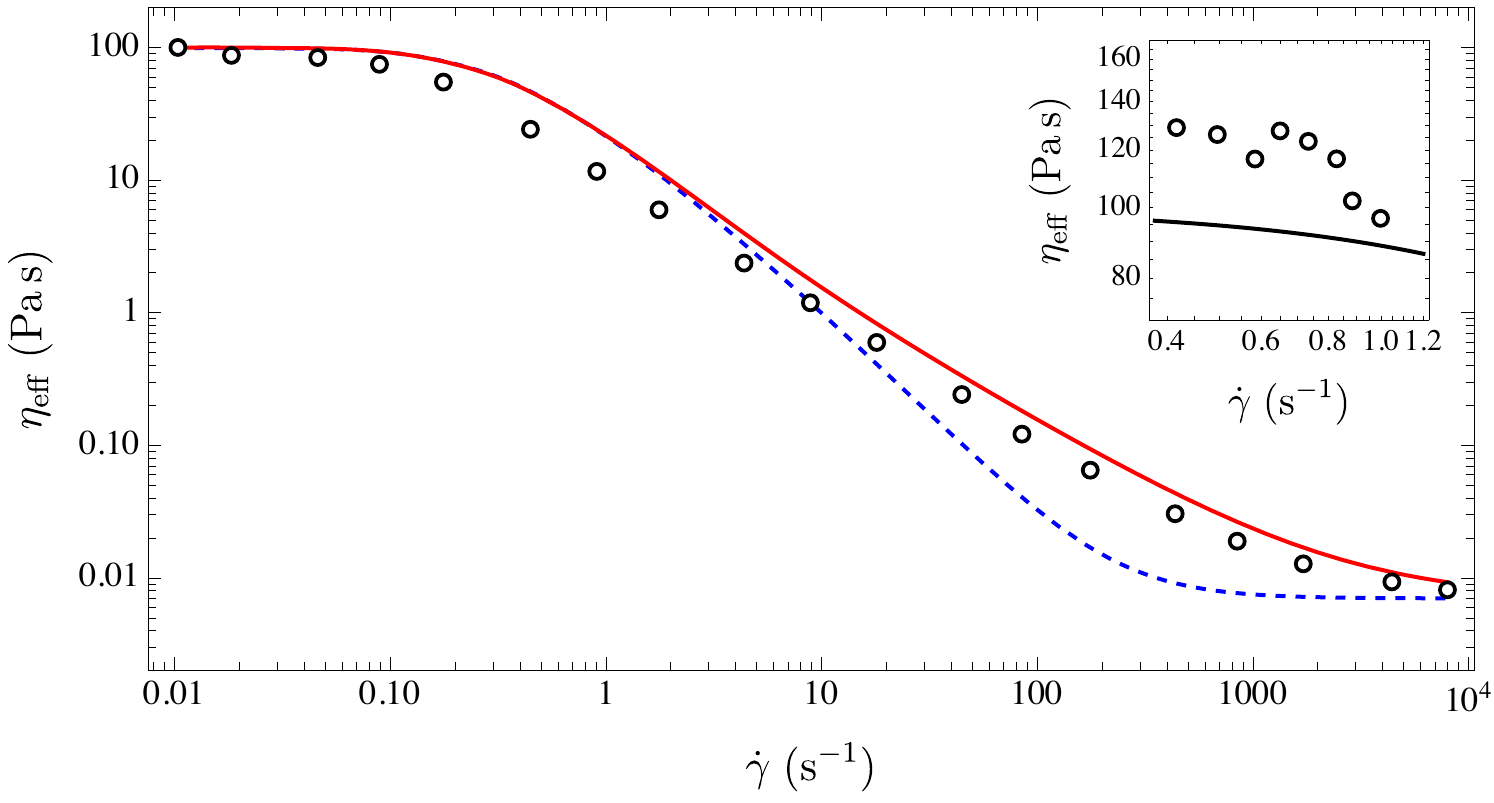}
\caption{\emph{With a rate-dependent relaxation time extrapolated from SAOS experiments we obtain an improved prediction of the steady shear viscosity.} We substitute $\omega$ with $\gd$ in the expression for $\taur$ used in Figure~\ref{fig:saosWLM}, with identical parameters, and find the red solid curve as opposed to the blue dashed one, given by our model with constant $\taur$. In the inset, we show the prediction with the same rate-dependent $\taur$ of the effective viscosity in extensional flows (black solid line), that captures the trend of the experimental data (circles) in the available window of deformation rates. The experimental data are by Haward \& McKinley~\cite{Haward_2012} and were rescaled in the case of planar extension to match our choice of rate and viscosity.
}\label{fig:steadyWLM}
\end{figure*}

Of course, many classical models are in principle amenable of such a generalization, nevertheless it is significant to check whether the rate dependence postulated from SAOS data can be successfully applied to steady shear flows as well. To this end, we exploit the important property that, in our framework, $\log\Bel$ depends only on $\Wi$ in simple shear flows and use the data presented in Figure~\ref{fig:steadystate} to obtain a numerical interpolation of the dependence of $\etaeff/\kappa$ and $\psieff/\kappa$ on $\Wi$. With these master curves we can retrieve the rheological predictions for any value of the parameters $\kappa$, $\eta$, and $\taur$, even when $\taur$ depends on $\gd$. The curve that we obtain by assuming relation \eqref{eq:taur_down}, with parameters fitted on the SAOS data and $\gd$ in place of $\omega$, is already much closer to the experimental behaviour than what we obtain with constant parameters (Figure~\ref{fig:steadyWLM}). Comparing the same model with extensional flow data (inset of Figure~\ref{fig:steadyWLM}) we see that it captures the mild thinning behaviour, but the viscosity values are still about 30\% off.

It is possible to indicate a general strategy to test whether an $\omega$-dependent $\taur$ is sufficient to reproduce the measured viscoelastic moduli. If the relations \eqref{eq:G1}--\eqref{eq:G2} can describe a given material, one can deduce that
\begin{equation}
\taur(\omega)=\frac{G'(\omega)}{\omega G''(\omega)-\omega^2\eta}.
\end{equation}
We can then extract this functional form of the relaxation time from the data and then try and see if, with such $\taur(\omega)$, we can fit the experimental values of $G'$ and $G''$. If the fitting is not satisfactory, we must conclude that \eqref{eq:G1} and \eqref{eq:G2} are not enough to represent the linear viscoelastic spectrum.
For instance, we expect difficulties in reproducing data for materials that feature a broad spectrum of relaxation times and non-constant elastic parameter $\kappa$, such as low density polyethylene melts \cite{Meissner_1975}, which are usually fitted with multi-modal equations \cite{Ferras_2019}.

\section{A note on dimensionless numbers}\label{sec:dimensionless}

In our presentation we preferred to make use of dimensional parameters for an easier link with measurable quantities. Nevertheless, it is important to identify dimensionless parameters that can help understanding the scaling behaviour of our model. The most common dimensionless numbers related to viscoelastic fluids are the Deborah ($\De$) and Weissenberg ($\Wi$) numbers mentioned in the previous section (see the note by Dealy~\cite{Dealy_2010} for a clear discussion of their definition). Clearly, they both compare the relaxation time $\taur$ with a characteristic rate associated with the fluid flow and remain relevant also in our model.
In fact, $\De=1$ and $\Wi=1$ mark important transition points in the frequency- and rate-dependence of the material functions, respectively. This can be seen from Figures~\ref{fig:saos} and \ref{fig:steadystate}, where the choice $\taur=1$ s allows to view the plots as if the appropriate dimensionless numbers were on the horizontal axis.

Our model, in comparison with that of a Newtonian fluid, features two extra parameters, $\taur$ and $\kappa$. Given that the former leads to $\De$ or $\Wi$, depending on the flow, we clearly need to introduce another dimensionless group. 
To quantify the overall relative importance between the plastic dissipation due to elasticity and viscous forces (with no reference to a specific flow)
we can consider the quantity $(\kappa\taur+\eta)/\eta$.
This number is very useful in two ways: first, it describes the  (log-scale) range within which the effective viscosity can vary; second, it represents the relative importance of plasticity with respect to viscosity in the material response observed at low values of $\De$ in oscillatory flows.  
At high values of $\De=\omega\taur$, the dissipation is mostly viscous, plasticity does not intervene, and the material response is that of a viscoelastic solid.

The situation in flows with constant strain rate is more involved. First of all, we should stress that using the ratio of the first normal stress difference over the shear stress to measure the degree of elasticity or nonlinearity of the fluid is misleading.
In fact, such a ratio becomes 
$\gd\psieff/\etaeff$, a quantity that is identically zero in extensional flows, irrespective of the presence of elastic stresses.
We then consider the definition $\Wi=\gd\taur$ (independent of the flow type) and observe that, in extensional flows, $\Wi$ is hardly relevant, since the stress growth and steady state are rate-independent. 
In simple shear, we can conclude that for large $\Wi$ the behaviour of the fluid is essentially viscous. This is due to the fact that a faster rotation of the principal directions of the elastic stress makes relaxation more efficient, leading to smaller values of the elastic strain and, consequently, stress.

\section{Conclusions}\label{sec:conclusions}

We introduced a class of tensorial models aimed at describing viscoelastic materials. The cornerstones of this framework are an elastic stress that depends logarithmically on a suitable measure of strain and the choice of letting the elastic strain evolution emerge from two distinct evolution equations, one for the current deformation and the other for a tensorial descriptor of the elastically-relaxed state. 
While the former is a necessary kinematic relation between velocity and deformation, the latter involves constitutive choices that are based on arguments borrowed from solid plasticity. Our line of thought differs considerably form the classical Oldroyd's approach. 
Even though we can derive an equation for a quantity akin to a conformation tensor, the objective rate entering its evolution is not a matter of choice, as it descends directly from the kinematic evolution of the current deformation gradient. 
We stress that, in our framework, viscoelastic fluids emerge as an interpolation between purely viscous fluids and solids, controlled by a relaxation time parameter ranging from zero (viscous fluid) to infinity (viscoelastic solid).

We have shown that a simple model with constant material parameters performs very well in reproducing the qualitative behaviour of viscoelastic fluids observed in rheometric experiments. Moreover, it helps understanding the origin of the difference in extensional and shear rheology and the relative importance of viscous, elastic, and plastic effects.
Another important feature of this model is that it avoids the erroneous prediction of an exponential growth of the elastic stress in extensional flows that sometimes arises in connection with elastic models of neo-Hookean type.
We explored the predictions of our constitutive model in basic viscometric flows but the tensorial nature of the model makes it applicable to general flows. A more extended analysis of rheological features that may appear in nonlinear and possibly non-homogeneous flows requires the development of reliable numerical schemes, such as stable finite-element approximations. This will be the subject of future work and will allow us to compare the model with experimental observations in large amplitude oscillatory flows and more complex pressure-driven flows through channels or pipes.

To approach the modelling of real fluids, it is important to consider the presence of multiple relaxation times. This can be done within our framework by letting the relaxation time parameter depend on other relevant quantities. We provide a first example of what can be achieved in this way by addressing a situation in which an abrupt change in the elastic response during uniaxial extension prevents the attainment of steady flows.
Our findings suggest the presence of a phenomenon that can be described as a progressive polymeric jamming, in which the relaxation time diverges due to the experiment geometry. 
We have also shown how to capture the rheological behaviour of wormlike micellar solutions by means of a rate-dependent relaxation time.
Meanwhile, we can indicate as the presence of multiple relaxation modes a challenging aspect of real fluids that may require important generalizations of our model.

How to fully harness the freedom in modelling the relaxation time parameter to capture further rheological behaviours, such as viscoplastic yielding and softening, and describe different classes of complex fluids will be the subject of future research. Nevertheless, we can foresee that the phenomenon of yielding could be reproduced by setting the relaxation time equal to a sufficiently large value for small elastic strains and then letting it decrease rather dramatically once a given threshold is overcome. Different choices of such a decrease should be motivated on the basis of microscopic models to achieve a better understanding of the origin of experimental observations.

Another intriguing question for future research is the inclusion of thixotropic effects. From our results, we cannot draw conclusions about the time dependence of the model parameters generated by adaptation effects or aging because we only considered an instantaneous dependence of $\taur$ on some characteristic time scale. Nevertheless, our modelling scheme can be extended to cover more complex behaviors.

\begin{acknowledgments}
Project funded by the European Union -- NextGenerationEU under the National Recovery and Resilience Plan (NRRP), Mission 4 Component 2 Investment 1.1 - Call PRIN 2022 No.\ 104 of February 2, 2022 of Italian Ministry of University and Research; Project 202249PF73 (subject area: PE - Physical Sciences and Engineering) ``Mathematical models for viscoelastic biological matter''.
During the development of this research, the work of G.G.G.\ was partially supported by the National Group for Mathematical Physics (GNFM) of the Italian National Institute for Advanced Mathematics (INdAM).
\medskip

\noindent\textbf{Authors' contributions}

G.G.\ Giusteri conceived the model and M.A.H Alrashdi performed analytical and numerical computations. Both authors edited and revised the manuscript, and gave final approval for publication.

\medskip
\noindent\textbf{Data Availability Statement}

The data that support the findings of this study are available upon reasonable request.
\end{acknowledgments}


\nocite{*}
\bibliography{relaxed_state_viscoelastic_model}

\end{document}